\documentclass[12pt]{article}
\usepackage{putex}
\begin{document}

\preprint{PUPT-2498}

\title{Segmented strings and the McMillan map}
\authors{Steven S. Gubser, Sarthak Parikh, and Przemek Witaszczyk\footnote{Also in the Institute of Physics, Jagiellonian University, S. Lojasiewicza 11, 30-348 Krakow, Poland}}
\institution{PU}{Joseph Henry Laboratories, Princeton University, Princeton, NJ 08544, USA}

\abstract{We present new exact solutions describing motions of closed segmented strings in $AdS_3$ in terms of elliptic functions.  The existence of analytic expressions is due to the integrability of the classical equations of motion, which in our examples reduce to instances of the McMillan map.  We also obtain a discrete evolution rule for the motion in $AdS_3$ of arbitrary bound states of fundamental strings and D1-branes in the test approximation.}

\date{February 2016}
\maketitle

\tableofcontents

\section{Introduction}
\label{INTRODUCTION}

Segmented strings, as introduced in \cite{Vegh:2015ska,Callebaut:2015fsa}, are strings whose shape at a given time is piecewise linear, or as close to piecewise linear as the ambient curved spacetime permits.  They generalize time-honored constructions \cite{Bardeen:1975gx,Artru:1979ye} which play a role in the Lund model of hadronization.  At first glance, segmented strings seem similar to a discretization of the string worldsheet as appropriate to an approximate numerical treatment of classical string motions (see \cite{Ishii:2015qmj} for interesting related work).  Indeed it was conjectured in \cite{Callebaut:2015fsa} that segmented string motions are dense in the set of all possible classical motions.  But segmented strings themselves involve no approximations.  Their purely algebraic formulation provides an interesting opportunity for exact results.

In \cite{Callebaut:2015fsa}, examples of closed segmented strings in $AdS_3$ were exhibited where at each time slice the shape of the string respected a particular dihedral symmetry.  The particular examples showed $D_3$ and $D_4$ symmetry.  $D_3$ and $D_4$ are the symmetry groups of an equilateral triangle and of a square, respectively, but in \cite{Callebaut:2015fsa} the initial shapes of the string were hexagonal in the $D_3$ case and octagonal in the $D_4$ case, and by extension these cases were referred to in \cite{Callebaut:2015fsa} as the regular hexagon and the regular octagon.

It was observed numerically in \cite{Callebaut:2015fsa} that the global times $\Delta\tau_i$ between kink collisions in the evolution of the $D_3$ and $D_4$ cases exhibited a form of quasi-periodicity.  This quasi-periodicity is suggestive of integrability, which should perhaps be expected since classical string motion in $AdS_3$ as controlled by the Nambu-Goto action is in fact integrable.  While considerations related to integrability helped inform concurrent treatments of related problems \cite{Vegh:2015ska,Vegh:2015yua}, it remained an outstanding problem to nail down precisely how to treat the regular polygons in a way that would make the quasi-periodicity manifest.  The first purpose of the current paper is to provide an analytic account of these motions.  It turns out that the evolution law for segmented strings simplifies in the presence of $D_n$ symmetry to an example of the so-called McMillan map \cite{Mcmillan1967some}, also sometimes referred to as the autonomous discrete Painlev\'{e} II equation.  Explicit solutions for the shape of the string and for the $\Delta\tau_i$ are available in terms of elliptic functions.

In \cite{Gubser:2016wno}, it was pointed out that the evolution of segmented strings become even simpler when one derives their equation of motion not from the Nambu-Goto action but instead the Wess-Zumino-Witten action appropriate for strings which are directly coupled to the three-form field strength which supports the $AdS_3$ geometry.  We will think of $AdS_3$ as supported entirely through Ramond-Ramond (RR) three-form field strength (as, for instance, in the $AdS_3 \times S^3 \times T^4$ near-horizon geometry of the D1-D5 bound state).  Then the Nambu-Goto action is appropriate for fundamental strings (f1), while the Wess-Zumino-Witten action is appropriate for D1-branes.  (We do not consider non-zero background values for the IIB axion.)  It was suggested in \cite{Gubser:2016wno} that one could go further and find a discrete evolution law for arbitrary D1-f1 bound states.  Our second purpose in this paper is to formulate that discrete evolution law and to show that it has the same algebraic property as the one for fundamental strings: namely, only addition, subtraction, multiplication, and division are required.  For fundamental strings, this implies that the locations of kink collisions can all be chosen consistently with coordinates in any desired number field, for example the rationals $\mathbb{Q}$ or an extension of the rationals by real algebraic numbers.  For the D1-f1 bound state, there is one additional parameter $\kappa$, which in general is irrational, so that the minimal field one can employ to describe the locations of kink collisions is the field extension $\mathbb{Q}[\kappa]$.

The organization of the rest of the paper is as follows.  In section~\ref{Action} we review the embedding of $AdS_3$ into $\mathbb{R}^{2,2}$ and the string actions that we will need in subsequent sections.  In section~\ref{Diamond} we show within a single causal diamond of a segmented string worldsheet that the equations of motion following from the actions of section~\ref{Action} can be satisfied using a simple interpolation ansatz from which the final form of the discrete evolution rule can be extracted.  In section~\ref{McMillan}, we specialize to $D_n$-symmetric configurations and show how the evolution laws of section~\ref{Diamond} reduce to instances of the McMillan map.  We then explain the solution in terms of elliptic functions and provide an explicit example to compare with \cite{Callebaut:2015fsa}.  We finish with a brief discussion in section~\ref{Discussion}.

While this work was nearing completion, we received \cite{Vegh:2016hwq}, which demonstrates that general motions of segmented fundamental strings are integrable in terms of an affine Toda lattice.  This beautiful formalism must be capable of reproducing our results for the f1 case, and perhaps it can be extended to arbitrary D1-f1 bound states.

\section{Global coordinates and string actions}
\label{Action}

We consider $AdS_3$ to be the hyperboloid $X^2 = -1$ in $\mathbb{R}^{2,2}$, where
 \eqn{Xnorm}{
  X^2 = X^\mu X_\mu = \eta_{\mu\nu} X^\mu X^\nu = -u^2 - v^2 + x^2 + y^2
   \qquad\hbox{where}\qquad
    \eta_{\mu\nu} = \diag\{-1,-1,1,1\} \,,
 }
and we use $X^\mu = (u,v,x,y)$ when convenient so as to avoid the possibility of confusing $X^2 = X^\mu X_\mu$ with the $2$ component of $X^\mu$.  The hyperboloid $X^2 = -1$ contains closed timelike curves which are removed by passing to the global cover.  All the considerations of the current work survive when we pass to the global cover, so we will focus just on the hyperboloid.  To see that the closed timelike curves on the hyperboloid cannot be a problem for segmented strings, an important point is that an arbitrary causal diamond in the bulk spans a global time $\Delta\tau$ less than $\pi$, which is the light-crossing time of $AdS_3$.

The proper starting point for D1-f1 bound states is the Dirac-Born-Infeld plus Wess-Zumino action, appearing for example in \cite{Douglas:1995bn}.  For our purposes a more informal treatment suffices.  Namely, consider the action
 \eqn{ModifiedPolyakov}{
  S = -{\tau_1 \over 2} \int d^2 \sigma \,  
    \partial_a Y^M \partial_b Y^M 
    (\sqrt{-h} h^{ab} G_{MN} + \kappa \epsilon^{ab} B_{MN}) \,,
 }
where $\tau_1$ is the total tension and $\kappa \in (0,1)$ is a parameter that controls how strongly the string couples to the RR two-form potential $B_{MN}$.  (We will not have occasion to refer to the Neveu-Schwarz two-form potential explicitly, so our use of $B_{MN}$ for the RR two-form potential should not cause confusion.)  The $Y^M$ are coordinates on $AdS_3$ rather than $\mathbb{R}^{2,2}$.  The equation of motion arising from \eno{ModifiedPolyakov} is 
 \eqn{MPeom}{
  \partial_a \sqrt{-h} h^{ab} \partial_b Y^M + 
    \sqrt{-h} h^{ab} \Gamma^M_{NL} \partial_a Y^N \partial_b Y^L - 
    {\kappa \over 2} \epsilon^{ab} H^M{}_{NL} \partial_a Y^N \partial_b Y^L 
   = 0\,.
 }
To give a more efficient presentation of these equations, let's choose coordinates $(\sigma^0,\sigma^1)$ on the worldsheet such that $\sqrt{-h} h^{ab} = \eta^{ab} = \diag\{-1,1\}$ and $\epsilon^{01} = -1$.  The first two terms in \eno{MPeom} may be replaced by $\partial_a \partial^a X_\mu - (\partial_a X^\rho \partial^a X_\rho) X_\mu$, where the second term comes from a Lagrange multiplier treatment enforcing the constraint $X^2 = -1$, and no additional connection term is needed because $\eta_{\mu\nu}$ is a flat metric.  To handle the third term, we set $H_{\mu\nu\lambda} = {1 \over 2} \epsilon_{\mu\nu\lambda\rho} X^\rho$ on $\mathbb{R}^{2,2}$, where $\epsilon_{-1,0,1,2,} = 1$.  The restriction of $H_{\mu\nu\lambda}$ to the $AdS_3$ hyperboloid is indeed the $SO(2,2)$-invariant volume form.\footnote{A slight subtlety is that for general $H_3$ on $AdS_3$, the lagrange multiplier term would pick up a contribution proportional to $H_3$.  For the $SO(2,2)$-invariant volume form, this term vanishes.  Equivalently, one can check that the evolution \eno{MPreformed} preserves the condition $X^2 = -1$.}  Then the equation of motion \eno{MPeom} becomes
 \eqn{MPreformed}{
  \partial_a \partial^a X_\mu - (\partial_a X^\rho \partial^a X_\rho) X_\mu - 
    \kappa \epsilon^{ab} \epsilon_{\mu\nu\lambda\rho} X^\nu 
     \partial_a X^\lambda \partial_b X^\rho = 0 \,.
 }
We will prefer to work with lightcone coordinates and the corresponding derivatives:
 \eqn{SigmaPlusMinus}{
  \sigma^\pm = {1 \over 2} (\sigma^0 \pm \sigma^1) \qquad\qquad
  \partial_\pm = \partial_0 \pm \partial_1 \,.
 }
Then \eno{MPreformed} becomes
 \eqn{MPlightcone}{
  \partial_+ \partial_- X_\mu - (\partial_+ X^\rho \partial_- X_\rho) X_\mu +
    \kappa \epsilon_{\mu\nu\lambda\rho} X^\nu 
     \partial_+ X^\lambda \partial_- X^\rho = 0 \,.
 }
These equations of motion are supplemented by the constraints
 \eqn{MPconstraints}{
  (\partial_+ X^\rho)^2 = (\partial_- X^\rho)^2 = 0 \,.
 }
The equivalent of these constraints in terms of the $Y^M$ variables can be derived as usual by varying \eno{ModifiedPolyakov} with respect to $h_{ab}$.

\section{An interpolation ansatz}
\label{Diamond}

The idea of segmented strings is to build up the worldsheet as a mosaic of causal diamonds, each bordering four others along edges which are null.  In all examples developed so far, the edges are not just null paths in spacetime; they are in fact null geodesics.  Let $X = X(\sigma^-,\sigma^+)$ be the embedding function of the string into spacetime, and let each causal diamond be the restriction of $X(\sigma^-,\sigma^+)$ to a region $(\sigma^-,\sigma^+) \in (i,i+1) \times (j,j+1)$ for some $i,j \in \mathbb{Z}$.  For brevity we will use the notation $X(i,j) = X_{ij}$.  We will focus mostly on the causal diamond corresponding to $i=j=0$.  Its corners in spacetime are $X_{00}$, $X_{10}$, $X_{01}$, and $X_{11}$.  In order to evolve the segmented string forward in time, we have to be able to take $X_{00}$, $X_{10}$, and $X_{01}$ as initial data and produce $X_{11}$.  

More precisely, the ``forward null triple'' $(X_{00},X_{10},X_{01})$ has to satisfy the constraints
 \eqn{DeltaConstraints}{
  \Delta_L^2 = \Delta_R^2 = 0 \qquad\hbox{where}\qquad
   \Delta_L = X_{10} - X_{00} \quad\hbox{and}\quad
   \Delta_R = X_{01} - X_{00} \,.
 }
In \eno{DeltaConstraints} and below, we have in mind that all the $X_{ij}$ are points in $\mathbb{R}^{2,2}$ which lie on the $AdS_3$ hyperboloid, while $\Delta_L$ and $\Delta_R$ are null vectors in $\mathbb{R}^{2,2}$.  We require further that the null vectors $\Delta_L$ and $\Delta_R$ must be forward-directed, so that $X_{10}$ and $X_{01}$ are at a later global $AdS_3$ time than $X_{00}$.\footnote{Because a causal diamond never stretches over a larger global time interval than $\Delta\tau = \pi$, we do not need to concern ourselves with the presence of closed timelike curves on the hyperboloid.}  The evolution law must take as input a forward null triple and deliver a backward null triple $(X_{11},X_{10},X_{01})$ (with the same $X_{10}$ and $X_{01}$).  We require as part of the definition of a backward null triple that $X_{11} - X_{10}$ and $X_{11} - X_{01}$ are forward-directed null vectors.  We can think of the process of mapping a forward null triple $(X_{00},X_{10},X_{01})$ into a backward null triple $(X_{11},X_{10},X_{01})$ as tantamount to adding the causal diamond with corners $X_{00}$, $X_{10}$, $X_{01}$, and $X_{11}$ to the worldsheet.  In \cite{Gubser:2016wno} a more complete account was given of how local moves of this type can be used to build up a complete worldsheet, in particular a closed worldsheet in which we require $X_{i-N_L,i+N_R} = X_{ij}$ for all $i$ and $j$.  Briefly, in lieu of specifying Cauchy data on a spatial slice across the worldsheet, one specifies the $X_{ij}$ along a ``serrated slice'' $S$ which runs from left to right across the string so that if $ij \in S$, the next point to the right is either $i-1,j$ or $i,j+1$.

The discussion of the previous paragraphs leaves open the question, how precisely do we determine $X_{11}$ given $X_{00}$, $\Delta_L$, and $\Delta_R$?  When the aim is to find exact solutions of a particular string equation of motion, say \eno{MPlightcone}, we can provide an answer as follows.  Let's prescribe as initial conditions
 \eqn{Xinit}{\seqalign{\span\TL & \span\TR & \qquad\span\TT}{
  X(\sigma^-,0) &= X_{00} + \sigma^- \Delta_L & for $\sigma^- \in (0,1)$  \cr
  X(0,\sigma^+) &= X_{00} + \sigma^+ \Delta_R & for $\sigma^+ \in (0,1)$\,,
 }}
and ask what $X(\sigma^-,\sigma^+)$ satisfies \eno{MPlightcone} subject to \eno{Xinit}.  We expect $X_{11}(\sigma^-,\sigma^+) = X_{00} + \sigma^- \Delta_L + \sigma^+ \Delta_R + \hbox{(non-linear corrections)}$, where the non-linear corrections are quadratic or higher order in $\sigma^\pm$.  It is easily checked that the following ansatz works:
 \eqn{Xansatz}{
  X(\sigma^-,\sigma^+) = 
    {1 + (1+\kappa^2) \sigma^- \sigma^+ \Delta_L \cdot \Delta_R / 2 \over
     1 - (1-\kappa^2) \sigma^- \sigma^+ \Delta_L \cdot \Delta_R / 2} X_{00} + 
    {\sigma^- \Delta_L + \sigma^+ \Delta_R + \kappa \sigma^+ \sigma^- N \over
     1 - (1-\kappa^2) \sigma^- \sigma^+ \Delta_L \cdot \Delta_R / 2} \,,
 }
where $\Delta_L \cdot \Delta_R = \eta_{\mu\nu} \Delta_L^\mu \Delta_R^\nu$ and
 \eqn{Ndef}{
  N^\mu = \epsilon^\mu{}_{\nu\lambda\rho} X_{00}^\nu 
   \Delta_L^\lambda \Delta_R^\rho \,.
 }
Specializing to $\sigma^+ = \sigma^- = 1$, we obtain
 \eqn{Xoneone}{
  X_{11} = {1 + (1+\kappa^2) \Delta_L \cdot \Delta_R / 2 \over
     1 - (1-\kappa^2) \Delta_L \cdot \Delta_R / 2} X_{00} + 
    {\Delta_L + \Delta_R + \kappa N \over
     1 - (1-\kappa^2) \Delta_L \cdot \Delta_R / 2} \,.
 }
It is also easily checked that $X_{11} - X_{10}$ and $X_{11} - X_{01}$ are forward-directed null vectors, and that the global time elapsed between $X_{11}$ and $X_{00}$ is less than $\pi$.

Although the differential equation \eno{MPlightcone} is second order overall, it is first order in $\partial_-$ and first order in $\partial_+$.  As a result, initial data of the form \eno{Xinit} uniquely fixes a solution within the causal diamond we are focused on.  Thus the discrete evolution rule \eno{Xoneone} is the only choice we can make consistent with our general strategy.

\subsection{Properties of the interpolation ansatz}

The solution \eno{Xansatz} describing the motion of D1-f1 bound states elegantly interpolates between two well-known extremes.  The action \eno{ModifiedPolyakov} and following from it equation \eno{MPlightcone} are parametrized by the continuous RR potential coupling $\kappa$. It is straightforward to see that the two limiting values $\kappa\to 0$ and $\kappa\to 1$ turn \eno{MPlightcone} respectively into the conventional Nambu-Goto equation for the f1 string solely, or to the WZW evolution for the D1-brane.  In both cases the solution is provided by an appropriate limit of \eno{Xansatz}.

It helps our intuition to examine a particularly simple kinematic setup. With the aid on the underlying $SO(2,2)$ symmetry we can choose the future null triple $(X_{10},X_{00},X_{01})$ to be
 \eqn{forwardTriple}{
 X_{00} = \begin{pmatrix} 1 \cr 0\cr 0 \cr 0 \end{pmatrix},\quad
 X_{10} = \begin{pmatrix} 1 \cr a\cr -a\cr 0 \end{pmatrix},\quad {\rm and\ } \quad
 X_{01} = \begin{pmatrix} 1 \cr b\cr b\cr 0 \end{pmatrix}\,,
 }
where $a$ and $b$ are positive.  Then according to \eno{Xoneone},
 \eqn{X11kappa}{
X_{11}(\kappa) = {1 \over 1+(1-\kappa^2)ab} \begin{pmatrix}
1-(1+\kappa^2)ab \cr a+b \cr -a+b  \cr -2 ab \kappa
\end{pmatrix} \,.
 }
Comparing to (37) and (50) of \cite{Gubser:2016wno}, we see that $X_{11}(\kappa)$ indeed smoothly interpolates between the f1 case ($\kappa=0$) and the D1 case ($\kappa=1$).  Furthermore, if we demand that $X_{11}$ lies at a later time than both $X_{10}$ and $X_{01}$, in the sense of the usual time ordering on the $AdS_3$ hyperboloid, then we get (recalling $a,b > 0$)
 \eqn{KappaLimits}{
  \kappa^2 < 1 + {1 \over ab} \,.
 }
Now, $a$ and $b$ are arbitrary positive numbers; in particular, they can be made large.  So we conclude on causal grounds alone that we must choose $\kappa \in [-1,1]$.

The set of points on $AdS_3$ which are null separated from both $X_{10}$ and $X_{01}$ is a one-dimensional locus, and \eno{X11kappa} provides a convenient parametrization of it.  One might therefore wonder if \eno{Xoneone} is the only possible evolution law consistent with $SO(2,2)$ symmetry and the requirement of propagating consistent forward null triples into consistent backward null triples.  (Consistent here refers to the requirement that forward null triples must satisfy the null constraints \eno{DeltaConstraints}, and the analogous requirement for backward null triples.)  In fact, \eno{X11kappa} is not unique in this regard: for example, we could use instead
\eqn{Alternative-X11}{ 
 X_{11}^{\rm alternative} &= X_{00} + \frac{1}{1-\Delta_L\cdot\Delta_R/2} \Bigg[ 
   \left( 1-\frac{\kappa^2}{2}\Delta_L\cdot\Delta_R \right) 
    (\Delta_L + \Delta_R + (\Delta_L \cdot \Delta_R) X_{00})  \cr
 &\qquad\qquad{} + 
  \kappa \sqrt{\left( 1 - \frac{1}{2}\Delta_L \cdot \Delta_R \right) 
    \left( 1 - \frac{\kappa^2}{2}\Delta_L \cdot \Delta_R \right)} N \Bigg]  \cr
 &= {1 \over 1 + ab} 
  \begin{pmatrix} 1 - ab (1+2ab\kappa^2) \\
   (a+b) (1+ab\kappa^2) \\ (-a+b) (1+ab\kappa^2) \\
   -2 ab\kappa \sqrt{(1+ab)(1+ab\kappa^2)} \end{pmatrix} \,.
 }
It may be, however, that \eno{Xoneone} is essentially the only possible evolution law which is purely rational---that is, involving no square roots.

\section{The McMillan map}
\label{McMillan}

\subsection{Regular polygons}
To see an explicit example in which the language of serrated slices greatly simplifies the description of classical string motion, let us apply the evolution law from section \ref{Diamond} to closed strings in the shape of regular polygons.  More precisely, in this section we describe the motion of regular $2n$-gons by tracking the collisions of the $2n$ kinks which constitute the segmented string.  The kinks travel along null geodesic trajectories, and their collision points $X_{ij}$ obey the evolution law \eno{Xoneone}.

To make the symmetries of the regular polygon manifest, let us describe the $AdS_3$ hyperboloid in terms of complex coordinates $(a,b)$ related to the embedding space and the global coordinates as 
 \eqn{abDef}{
  a = u + iv = e^{i\tau} \cosh\rho \qquad\qquad b = x + iy = e^{i\phi} \sinh\rho \,.
 }
 We will assume invariance under certain discrete transformations in the $x$-$y$ plane. The simplest possibilities are cyclic symmetry $\mathbb{Z}_n$ and dihedral symmetry $D_n$.  To be explicit, let's introduce
 \eqn{omegaDef}{
  \omega = e^{\pi i / n}
 }
so that $\omega^{2n} = 1$.  Multiplying $b$ by $\omega$ generates ${\mathbb Z}_{2n}$ rotation, so to get ${\mathbb Z}_n$ invariance we demand invariance under
 \eqn{cyclicSymmetry}{
  b \to \omega^2 b \,.
 }
To impose dihedral symmetry we additionally require invariance under
 \eqn{bConjugation}{
  b \to \bar{b} \,,
 }
where bars mean complex conjugation.

In \cite{Callebaut:2015fsa}, initial conditions were specified at $\tau=0$ in terms of the corners $H_i^\mu$ of a $2n$-gon and initial forward-directed null velocities $V_i^\mu$ of those corners. To get a serrated slice instead of initial data at $\tau=0$, we propagate forward and backward ``half a step.''  Focusing on three neighboring corners (kinks), we form
 \eqn{TwoX}{
  X_{00} &= H_1 - \lambda V_1 = H_{2n} - \lambda V_{2n}  \cr
  X_{01} &= H_1 + \lambda V_1 = H_2 + \lambda V_2 \,,
 }
where $\lambda > 0$ is chosen so that the second equalities in both lines of \eno{TwoX} hold. This works assuming that the angle between the spatial parts of $V_1$ and $V_{2n}$ is obtuse, so that kinks $1$ and $2n$ are separating as we move forward in time.  By construction, the points $X_{ij}$ give the space-time coordinates of the collision sites of kinks in the embedding space, whereas the indices $(i,j)$ label the location of collisions on the worldsheet lattice, introduced in \cite{Gubser:2016wno}. Note that in \eno{TwoX} the point $X_{00}$ lies in the past of $\tau = 0$ while $X_{01}$ lies in its future.  Let's assume that the original $H_n$ are numbered in order of increasing phase of their $b$ components.  Then on account of cyclic symmetry, we can write the serrated slice centered on $\tau = 0$ as
 \eqn{CyclicSerratedSlice}{
  X_{-k,k} &= (a_0,\omega^{2k} b_0)  \cr
  X_{1-k,k} &= (a_1,\omega^{2k} b_1) \,,
 }
where $k$ runs from $0$ to $n-1$.  It will be recalled from section~\ref{Action} that a periodic serrated slice has $X_{i-N_L,j+N_R} = X_{ij}$ for all $i$ and $j$.  Here $N_L = N_R = n$.  Essentially, $k$ stands in lieu of the spatial coordinate $\sigma^1$ on the worldsheet.  The full set of collision points must take the form
 \eqn{CyclicAnsatz}{
  X_{m-k,k} = (a_m,\omega^{2k} b_m) \,,
 }
where $m$ stands approximately in lieu of a time coordinate on the worldsheet.  (Actually, $m$ is more closely related to $\sigma^-$, but the distinction between $\sigma^0$ and $\sigma^-$ is not important in this context.)  We want a recursion relation that will determine $(a_{m+2},b_{m+2})$ in terms of $(a_m,b_m)$ and $(a_{m+1},b_{m+1})$. Of particular interest in \cite{Callebaut:2015fsa} was the sequence $\tau_m$ of arguments of $a_m$, defined so that $\Delta\tau_m = \tau_m-\tau_{m-1}$ are positive.  It was found that $\Delta\tau_m$ shows quasi-periodic behavior when plotted against $\tau_m$. 

In the rest of this section we write down the desired recursion relations, show that the classical motion of such polygonal strings is integrable, and write down explicit expressions which solve the recursion relations and describe the motion of an arbitrary bound state of f1 strings and D1-branes. We then specialize to the pure f1 string, and the pure D1-brane. Finally we compare our analytic expressions for the f1 string with the numerical results of \cite{Callebaut:2015fsa} and find perfect agreement.

\subsection{Recursion relations}

The usefulness of imposing the $D_n$-symmetry is that it allows us to assume
 \eqn{Bassume}{
  b_m = \omega^{-m} B_m \qquad\hbox{where}\qquad B_m \in {\mathbb R} \,.
 }
Substituting the ansatz \eno{CyclicAnsatz} in the evolution equation \eno{Xoneone}, with $b_m$ given by \eno{Bassume}, and using the constraints that kinks must travel along null trajectories, one obtains the following recursion relations:
 \eqn{Brecursion}{
  B_{m+2} - {2\mu_m B_{m+1} \over 
    1 + (1-\kappa^2) (B_{m+1} \Im \omega)^2} + B_m &= 0  \cr
  a_{m+2} - {2\beta_m a_{m+1} \over 
    1 + (1-\kappa^2) (B_{m+1} \Im \omega)^2} + a_m &= 0 \,.
 }
Here
 \eqn{mumForm}{
  \mu_m = (1-\kappa) \Re\omega + 
   \kappa (\Re\{ a_m \bar{a}_{m+1} \omega \} - B_m B_{m+1}) + 
   \kappa (1-\kappa) B_m B_{m+1} (\Im\omega)^2 \,,
 }
and a similar expression can be given for $\beta_m$ whose form is unenlightening.  For $\kappa=0$, it is obvious that $\mu_m$ does not depend on $m$.  For $\kappa \neq 0$, a straightforward but tedious calculation leads to $\mu_{m+1} - \mu_m = 0$, so that $\mu_m$ is still constant.  Therefore we will replace $\mu_m \to \mu$ where the quantity
 \eqn{muForm}{
  \mu = (1-\kappa) \Re\omega + 
   \kappa (\Re\{ a_0 \bar{a}_1 \omega \} - B_0 B_1) + 
   \kappa (1-\kappa) B_0 B_1 (\Im\omega)^2
 }
can be determined in terms of initial conditions.  This is a crucial step because once we substitute $\mu_m \to \mu$ in \eno{Brecursion}, the recursion relation for the $B_m$ decouples from the $a_m$, becoming a single non-linear second-order real difference equation.  Furthermore, the form of this difference equation is essentially an example of the  McMillan map \cite{Mcmillan1967some}, whose integrability has been well-studied in the literature.  As we will explain in the next section, closed form expressions for $B_m$ may be extracted in terms of elliptic functions.

Further study of the constraints leads to
 \eqn{aConstraint}{
 {a_{m+1} \over a_m} = {1 + B_m B_{m+1} \Re\omega + \sqrt{-B_m^2 - B_{m+1}^2 + 2B_m B_{m+1} \Re \omega - \left(B_m B_{m+1} \Im \omega\right)^2 }  \over 1 + B_m^2 }\,.
 }
There is no $\kappa$ dependence in \eno{aConstraint} because the equation of motion \eno{Xoneone} is not required for its derivation; instead one needs only $X_{m-k,k}^2 = -1$ and the null condition on kink trajectories.  Once the $B_m$ coefficients have been found, the constraint equation \eno{aConstraint} together with the initial conditions may be used to find the $a_m$ coefficients, bypassing the need to solve the more complicated recursion relation for the $a_m$ coefficients in the second line of \eno{Brecursion}.

The equations \eno{Brecursion} describe the motion of an arbitrary bound state of f1 strings and D1-branes, characterized by the value of $\kappa$. Before moving on to discuss the McMillan map in detail, let us note two special limits: 
\begin{itemize}
\item $\kappa = 0$:  corresponds to the motion of f1 strings described by the Nambu-Goto action. The recursion relations in this case simplify to
 \eqn{Brecursionf1}{
   B_{m+2} - {2 B_{m+1} \Re\omega \over 1 + \left( {B_{m+1}} 
    \Im\omega \right)^2} + B_m = 0 
  \qquad
  a_{m+2} - {2a_{m+1} \over 1 + \left( {B_{m+1}} \Im\omega \right)^2} + 
    a_m = 0\,.
 }
The procedure for solving this case is the same as for general $\kappa$: the $B_m$ obey a McMillan map, and the second order difference equation for the $a_m$, while relatively simple, is still less useful than the first order equation \eno{aConstraint}.  The only simplification is that $\mu = \Re\omega$ is fixed, independent of initial conditions.
\item $\kappa = 1$: corresponds to the motion of D1-branes described by the WZW action. In this case the McMillan map degenerates, and the recursion relations take a particularly simple form:
 \eqn{BrecursionWZW}{
 B_{m+2} - 2\mu B_{m+1} + B_m  = 0 \qquad \tilde a_{m+2} - 2\mu \tilde a_{m+1} + \tilde a_m  = 0  \,,
 }
where we have defined $\tilde a_m \equiv a_m \omega^{-m}$.  Note that in this case $\mu$ must be determined from the initial conditions.  Once this is done, \eno{BrecursionWZW} can be solved entirely in terms of ring operations.  Explicitly,
 \eqn{WZWSoln}{
 B_m &= c_1 \left( \mu - \sqrt{\mu^2-1} \right)^m + 
        c_2 \left( \mu + \sqrt{\mu^2-1} \right)^m \cr
 \tilde a_m &= d_1 \left( \mu - \sqrt{\mu^2-1} \right)^m + 
               d_2 \left( \mu + \sqrt{\mu^2-1} \right)^m\,,
 }
where $c_1$, $c_2$, $d_1$, and $d_2$ are constants (not all independent) determined from the initial conditions.
\end{itemize}

\subsection{Elliptic functions and the McMillan map}
Before we write down analytic expressions which solve the recursion relations \eno{Brecursion}, let us briefly review the McMillan map. During 1967-68 E.~M.~McMillan studied a class of discrete difference equations of the form~\cite{Mcmillan1967some}
 \eqn{McMapGeneral}{
 q_{m+2} + q_m = f(q_{m+1})
 } 
 with
 \eqn{fGeneral}{
 f(q_m) = -{ B q_m^2 + D q_m + E \over A q_m^2 + B q_m + C}\,,
 }
where $A, B, C, D, E$ are constants. A ``double quadratic'' function, given by
 \eqn{InvariantGeneral}{
 I_m \equiv A q_m^2 q_{m+1}^2 + B (q_m^2 q_{m+1} + q_m q_{m+1}^2) + C(q_m^2 + q_{m+1}^2) + D q_m q_{m+1}  + E (q_m + q_{m+1}) 
 } 
was argued to be invariant under the map, that is $I_m = I_{m+1}$ for all $m$.

The so-called McMillan map is a special case of \eno{fGeneral} corresponding to $A = 1, B=0, C=1, D = -2 \mu$ and $E=0$, so that
 \eqn{fMcMap}{
 f(q_m) = {2 \mu q_m \over 1 + q_m^2}\,,
 }
and the invariant under the map reduces to
 \eqn{McMapInvariant}{
 I_m = q_m^2 q_{m+1}^2 + q_m^2 + q_{m+1}^2 - 2 \mu q_m q_{m+1}\,,
 }
 where $\mu$ is a constant.
 Note that in this case the second order difference equation \eno{McMapGeneral} is also known in the literature as the autonomous form of the discrete Painlev\'{e}-II equation, although in a ``non-standard'' form -- in the ``standard'' form $q_m^2$ in the denominator in \eno{fMcMap} comes with a minus sign~\cite{Nijhoff1991similarity,Ramani1991discrete}.
Another case of interest to us (when considering the WZW $SL(2,{\bf R})$ evolution of D1-branes) corresponds to setting $A=0, B=0, C=1, D=-2\mu$ and $E=0$ in \eno{fGeneral}, so that
 \eqn{fWZW}{
 f(q_m) = 2\mu q_m\,,
 }
and the invariant in this case becomes
 \eqn{WZWInvariant}{
 I_m = q_m^2 + q_{m+1}^2 - 2\mu q_m q_{m+1}\,.  
 }
 
The McMillan map \eno{fMcMap} can be solved in terms of elliptic functions. The exact solution is given by
 \eqn{McMapSoln}{
  q_m = Q\; {\rm cn}(\Omega m + \phi, k_e)\,.
  }
  The four parameters $ Q, \Omega, \phi$ and $k_e$ are determined using four conditions -- two initial conditions, say at  $m=0, 1$, and the following two constraints,
 \eqn{McMapConstraints}{
  Q = k_e\frac{ {\rm sn}(\Omega, k_e)}{{\rm dn}(\Omega, k_e)} \qquad \mu = {{\rm cn}(\Omega, k_e) \over {\rm dn}^2(\Omega , k_e) }\,.
  }
  Here ${\rm cn}(u,k_e),\, {\rm sn}(u,k_e)$ and ${\rm dn}(u,k_e)$ are the Jacobi elliptic functions and $k_e$ is the elliptic modulus. Making use of the invariant \eno{McMapInvariant}, the constraints can be partially disentangled. The amplitude $Q$ and the elliptic modulus $k_e$ are found to be algebraic
  \eqn{QkeSoln}{
   Q^2 = { -1 + I_0 + \mu^2 + \sqrt{\left(-1+I_0 +\mu^2\right)^2+4I_0} \over 2} \qquad  k_e = \frac{|Q| \sqrt{1+  Q^2}}{\sqrt{\left(1+  Q^2\right)^2-\mu^2}}\,,
   }
  where $I_0$ is the McMillan invariant \eno{McMapInvariant}. 
 The frequency $\Omega$ and the phase $\phi$ can subsequently be determined using
 \eqn{OmegaphiSoln}{
 {\rm cn}\left(\Omega, k_e\right) =  {\mu \over 1+  Q^2} \qquad {\rm cn}\left(\phi, k_e\right) =  {q_0 \over Q}\,,
 }
where $q_0$ is handed as part of the initial data.

\subsection{Exact solution for string motion}
\label{ABSOLN}
We are now ready to write down the general solution of the $B_m$ recursion relation \eno{Brecursion} which takes precisely the form of a McMillan map. The solution is given by
 \eqn{BSoln}{
 B_m = { q_m  \over \sqrt{1-\kappa^2} \Im \omega} \,,
 }
 where $q_m$ is fleshed out in \eno{McMapSoln}, \eno{QkeSoln} and \eno{OmegaphiSoln}, and $B_0$ and $B_1$ are specified as initial conditions.
 The complex $a_m$ coefficients can be written as
 \eqn{}{
 a_m \equiv A_m \exp(i \tau_m)
 } 
 where the amplitude is given by
 \eqn{}{
 A_m =  \sqrt{1 + {q_m^2 \over (1-\kappa^2)(\Im \omega)^2}}  
 }
 and the phase is determined using the constraint relation \eno{aConstraint} which gives
 \eqn{DeltatauSoln}{
 \Delta \tau_m \equiv \tau_{m+1} - \tau_m = -i \log \left( { (1-\kappa^2)(\Im \omega)^2 + \sqrt{-J_m(\kappa)} \Im \omega +  q_m q_{m+1}  \Re \omega \over \sqrt{(1-\kappa^2) (\Im \omega)^2 + q_m^2} \sqrt{(1-\kappa^2) (\Im \omega)^2 + q_{m+1}^2}} \right)\,,
 }
 where 
 \eqn{JDef}{
 J_m(\kappa) \equiv q_m^2 q_{m+1}^2 + \left(1-\kappa^2\right) (q_m^2 + q_{m+1}^2 - 2  q_m q_{m+1} \Re \omega)
 }
  which yields
 \eqn{tauSoln}{
 \tau_m = \tau_0 + \sum_{k=0}^{m-1} \Delta \tau_k\,.
 }
 Here $\tau_0$ is provided as the third and final initial condition.  
 
 Note that when $\kappa = 0$, the equations describe the motion of fundamental strings governed by the Nambu-Goto action. In this limit $J_m(\kappa)$ reduces to the McMillan invariant \eno{McMapInvariant} since $\mu(\kappa = 0) = \Re \omega$. As we stated earlier, numerical results for the motion of a regular hexagon and an octagon controlled by the Nambu-Goto action were presented in \cite{Callebaut:2015fsa}; they find exact agreement with the analytic expressions \eno{BSoln}-\eno{tauSoln} written above. We briefly describe the example case of a regular hexagon in the next section.

\subsection{An example: the regular hexagon}

 We close this section by working out an explicit example. We will apply our analytic expressions to the case of a fundamental string ($\kappa =0$) in the shape of a regular hexagon ($n=3$) considered in \cite{Callebaut:2015fsa}. The initial conditions for the hexagon chosen in \cite{Callebaut:2015fsa} can be translated   in the language of serrated slices, to those for the $B_m$ and $a_m$ coefficients
 \eqn{InitialHex}{
 B_0 = 1 \qquad B_1 = 1 \qquad a_0 = \frac{1}{2} \left(\sqrt{7}-i\right)\,.
 }
 In terms of $q_m = B_m \Im \omega$ and $\tau_0$ the initial conditions are
 \eqn{InitialHexq}{
 q_0 = \sqrt{3}/2 \qquad q_1 = \sqrt{3}/2 \qquad \tau_0 = -\cot ^{-1}\left(\sqrt{7}\right)\,.
 }
 Using \eno{QkeSoln} we obtain
 \eqn{QkeHex}{
 Q = \sqrt{\frac{9+5 \sqrt{57}}{32}} \qquad k_e = \sqrt{\frac{1}{190} (95+3 \sqrt{57})}\,.
 }
 The frequency $\Omega$ and the phase $\phi$ are found by solving the equations \eno{OmegaphiSoln}
 \eqn{OmegaphiHex}{
 {\rm cn}(\Omega,k_e) = \frac{1}{16} (41-5 \sqrt{57}) \qquad {\rm cn}( \phi,k_e) = 2 \sqrt{\frac{6}{9+5 \sqrt{57}}}\,.
 }
 These equations may have more than one possible solutions. Care must be taken to ensure they are consistent with the following initial condition
 \eqn{ConsistencyOmegaphi}{
 {\rm cn}(\Omega + \phi) = {q_1 \over Q}\,.
 }
 A consistent pair of solutions for $\Omega$ and $\phi$ is
 \eqn{SambergHexphi}{
 \Omega &\approx 1.64294 + 4 d_1 K(k_e) + 4i d_2 K(\sqrt{1-k_e^2)}) \cr
 \phi &\approx 7.06205 + 4 c_1 K(k_e) + 4i c_2 K(\sqrt{1-k_e^2)})\,,
 }
 where $c_1, c_2, d_1, d_2$ are integers. As expected these solutions are arbitrary up to double periodicity, so any integral value for $c_1, c_2, d_1, d_2$ works. Here $K(k_e)$ is the complete elliptic integral of the first kind. The $q_m$ evolve according to $q_m = q_{\rm amp} {\rm cn}(\Omega m + \phi, k_e)$ where all the parameters have been computed above. The $q_m$ turn out to lie in the field extension ${\mathbb Q}[\sqrt{3}]$. The first few values for $q_m$ are
 \eqn{Hexqm}{
 \frac{\sqrt{3}}{2},\frac{\sqrt{3}}{2},-\frac{3 \sqrt{3}}{14} ,-\frac{307 \sqrt{3}}{446} ,-\frac{9631 \sqrt{3}}{137618},\frac{106748289 \sqrt{3}}{172349618},\frac{664717591693 \sqrt{3}}{1857024006686},\ldots
 } 
 The $B_m$ coefficients are then given by $B_m = q_m /\Im \omega$, and they turn out to lie in the rationals ${\mathbb Q}$. The first few values for $B_m$ are
 \eqn{HexBm}{
 1,1,-\frac{3}{7},-\frac{307}{223},-\frac{9631}{68809},\frac{106748289}{86174809},\frac{664717591693}{928512003343},-\frac{39977632574885123}{55400025215333527},\ldots
 }
 The $\Delta \tau_m$ can be computed using \eno{DeltatauSoln}. The first few values are
 \eqn{HexDeltatau}{
 \cot ^{-1}\left(\frac{3}{\sqrt{7}}\right),\tan ^{-1}\left(\frac{7 \sqrt{7}}{11}\right),\tan ^{-1}\left(\frac{1561 \sqrt{7}}{4043}\right),\tan ^{-1}\left(\frac{15344407 \sqrt{7}}{33645531}\right),\ldots
 }
 The $\Delta \tau_m$ found above can be used to find the individual $\tau_m$, corresponding to the $AdS_3$ global times at which the collision of kinks takes place. The first few values are
 \eqn{Hextau}{
 -\cot ^{-1}\left(\sqrt{7}\right),\cot ^{-1}\left(\sqrt{7}\right),\tan ^{-1}\left(\frac{15}{\sqrt{7}}\right),\pi -\tan ^{-1}\left(\frac{617}{167 \sqrt{7}}\right),\pi -\tan ^{-1}\left(\frac{9799}{52391 \sqrt{7}}\right),\ldots
 }
 Finally, the $a_m$ coefficients can now be determined. They belong to the field extension of the rationals to include both $i$ and $\sqrt{7}$, that is $\mathbb{Q}[i, \sqrt{7}]$. The first few values are
 \eqn{Hexa}{
 \frac{1}{2}  (\sqrt{7}-i),\frac{1}{2}  (\sqrt{7}+i),\frac{1}{14} (\sqrt{7}+15 i),\frac{1}{446} (-167 \sqrt{7}+617 i ),\frac{-52391 \sqrt{7}+9799 i}{137618},\ldots
 }
The $\Delta\tau_m$ as listed in \eno{HexDeltatau} agree with the results of \cite{Callebaut:2015fsa}.

\section{Discussion}
\label{Discussion}

In our current approach, the key to finding the evolution law \eno{Xoneone} for segmented strings is to explicitly analyze the shape of the string worldsheet inside a given causal diamond, say the one with corners $X_{00}$, $X_{10}$, $X_{01}$, and $X_{11}$.  We stipulate that the string worldsheet should coincide with a null geodesic on all four edges of the causal diamond.  This stipulation is crucial because it allows us to consistently join causal diamonds together.  Strikingly, the final form of the evolution law \eno{Xoneone} is purely algebraic, so simple that one might have guessed its form without the help of the interpolation ansatz \eno{Xansatz}.  We would like to understand more systematically how the class of generally covariant string actions maps onto the set of possible discrete evolution laws.  To pose a definite starting point: Is there an obvious classical action from which the alternative evolution law \eno{Alternative-X11} arises?

Our study of strings with $D_n$ symmetry is clearly the simplest of many possible problems one could consider, and already we encountered some interesting and non-trivial mathematics in the form of the McMillan map.  Our use of the McMillan invariant $I_m$ was limited to finding good ways of parametrizing initial conditions in terms of the parameters of the elliptic functions which solve the recursion relations.  It would be satisfying to assign some deeper physical meaning to $I_m$, and to extend this approach to integrability to less symmetrical string motions.

The distinctive feature of segmented strings is that once an evolution law like \eno{Xoneone} is established (or accepted), one does not require a continuous spacetime.  Instead, essentially algebraic constructions are enough, as outlined in \cite{Gubser:2016wno}.  We could for example replace $AdS_3$ by a level set of an indefinite quadratic form on a vector space $K^4$ where $K$ is a number field, provided the quadratic form has signature $(2,2)$ over the reals.  Integrable dynamics in such a ``spacetime'' might provide interesting new connections among dynamical systems, string theory, and number theory.

\section*{Acknowledgments}

This work was supported in part by the Department of Energy under Grant No.~DE-FG02-91ER40671. The work of PW was supported by the National Science Centre grant 2013/11/N/ST2/03812 and in part by 2012/06/A/ST2/00396.

\bibliographystyle{ssg}
\bibliography{regular}

\end{document}